\documentclass{moriond}

\usepackage[capitalise]{cleveref}

\def\Journal#1#2#3#4{{#1} {\bf #2}, #3 (#4)}

\def\PRL{\em Phys. Rev. Lett.}
\def\PRD{{\em Phys. Rev.} D}
\def\PRC{{\em Phys. Rev.} C}

\def\PRepD{\em Phys. Rept.}
\def\JPG{{\em J. Phys.}  G}
\def\EPJConf{\em EPJ Web Conf.}

\def\be{\begin{equation}}
\def\ee{\end{equation}}
\def\bea{\begin{eqnarray}}
\def\eea{\end{eqnarray}}

\newcommand{\Photo}{}

\begin{document}
\vspace*{4cm}
\title{Astroparticle Physics at the Forward Physics Facility}

\author{Dennis Soldin}

\address{Department of Physics and Astronomy, University of Utah,\\ Salt Lake City, UT 84112, USA}

\maketitle\abstracts{High-energy collisions at the High-Luminosity Large Hadron Collider (HL-LHC) will generate a substantial flux of particles along the beam collision axis that current LHC experiments cannot access. Multi-particle production in the far-forward region is particularly relevant to astroparticle physics. High-energy cosmic rays initiate extensive air showers (EASs) in the atmosphere, driven by hadron-ion collisions at low momentum transfer in the non-perturbative regime of QCD. Therefore, understanding high-energy hadronic interactions in the forward region is essential for interpreting EAS data and estimating backgrounds for astrophysical neutrino searches, among other applications. The Forward Physics Facility (FPF) is a proposal to construct a new underground cavern at the HL-LHC dedicated to hosting a range of experiments aimed at detecting particles beyond the current capabilities of LHC experiments. This article will provide an update on the current planning status of the FPF and emphasize its synergies with astroparticle physics.}

\section{Introduction}

\begin{figure}[tb]
    \vspace{-1.em}
    \centering
    \includegraphics[width=0.9\textwidth]{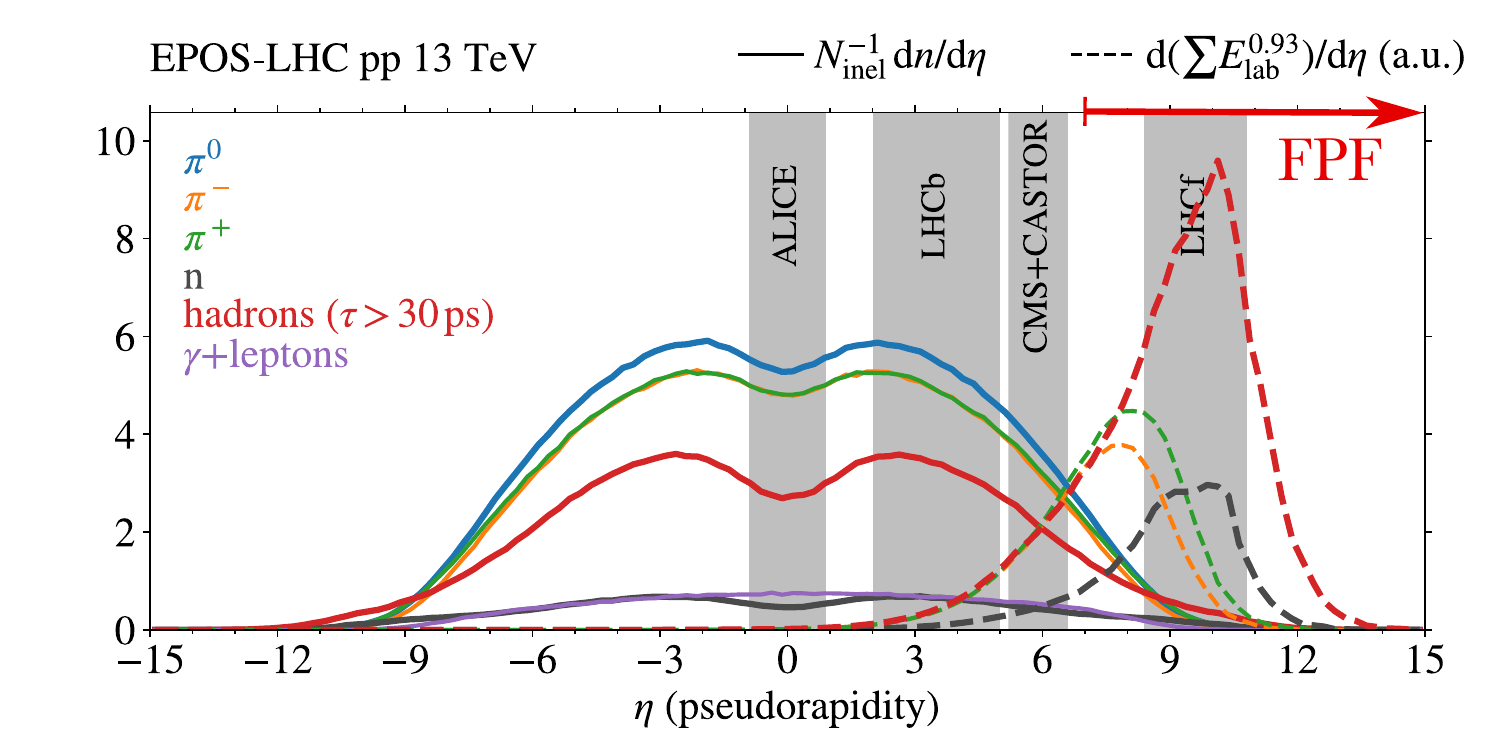}
    \vspace{-1.em}
    
    \caption{ \label{fig:eta-ranges}Simulated densities of particles~\protect\cite{MuonPuzzle} in arbritrary units (solid lines) in proton-proton collisions using EPOS-LHC. Dashed lines show the estimated number of muons produced by these particles, assuming an equivalent energy for the fixed target collisions in the laboratory frame, $E_{\rm{lab}}$, and $N_\mu \propto E_{\rm{lab}}^{0.93}$.
    }
    \vspace{-0.5em}
\end{figure}

Cosmic rays with energies above $10^{11}~\rm{GeV}$ enter Earth's atmosphere where they interact with air molecules, generating extensive air showers (EASs) that can be detected with large ground-based detector arrays. Determining the properties of initial cosmic rays, such as their energy and mass, relies on indirect measurements of particles detected at ground level by these arrays, which requires a simulation-based interpretation of the EAS development. The main challenge in the description of EASs lies in modeling hadronic interactions in the forward region over a wide range of energies\,\cite{Kampert,Engel}, which current collider facilities cannot directly probe. \Cref{fig:eta-ranges} depicts simulated particle densities from proton-proton collisions (solid lines), overlaid with the pseudorapidity ($\eta$) ranges relevant to current LHC experiments\,\cite{MuonPuzzle}. Dashed lines show estimated muon densities, $N_\mu$, produced by these particles, assuming $N_\mu \propto E_{\rm{lab}}^{0.93}$, where $E_{\rm{lab}}$ is the energy of secondary EAS particles in the laboratory frame. While mid-rapidity ranges minimally affect EAS particle production, the forward region ($\eta>4$) significantly influences the EAS development, which is driven by relativistic hadron-ion collisions in the atmosphere at low momentum transfer within the non-perturbative regime of quantum chromodynamics (QCD). Given the limitations of describing hadron production from first principles, and the lack of data from existing collider experiments, simulations highly rely on various phenomenological models of hadronic interactions, yielding substantial uncertainties. Hence, precise measurements of hadronic interactions under controlled experimental conditions, particularly in the far-forward region, are crucial for validating and refining existing EAS models.

%------------------------
%------------------------

\section{The Forward Physics Facility}

The Forward Physics Facility\,\cite{FPF_SP,FPF_WP} (FPF) is a proposal to build a new underground cavern at CERN, designed to accommodate a suite of experiments focused on the far-forward region during the high-luminosity LHC (HL-LHC) era. Current LHC detectors have uninstrumented areas along the beam line, potentially missing physics opportunities arising from the substantial flux of particles produced in the forward direction. Without the FPF, the HL-LHC would lack sensitivity to neutrinos and many particles proposed by various models beyond the Standard Model, including new force carriers, sterile neutrinos, and axion-like particles, for example. However, recent pathfinder experiments operating in the forward region at the LHC have successfully observed collider neutrinos directly for the first time and have demonstrated the potential for discovering new physics\,\cite{FASER1,SNDLHC1}. A diverse suite of experiments will be available at the FPF to exploit these physics opportunities by detecting neutrino interactions at the highest accelerator energies, advancing our understanding of particle interactions in the far-forward region.

The proposed site at CERN is situated at a depth of $88\,\rm{m}$ along the line-of-sight (LOS) of the ATLAS collision axis and is located $627\,\rm{m}$ west of the interaction point, shielded by over $200\,\rm{m}$ of rock. In an updated baseline layout, the facility will measure approximately $75\,\rm{m}$ in length and $11.8\,\rm{m}$ in internal width, providing the infrastructure needed to host a diverse set of experiments aimed at exploring various physics phenomena at pseudorapidities above $\eta\sim 7$.

\subsection{Experiments}

\begin{figure}[tb]
\centering
\includegraphics[width=1.0\textwidth]{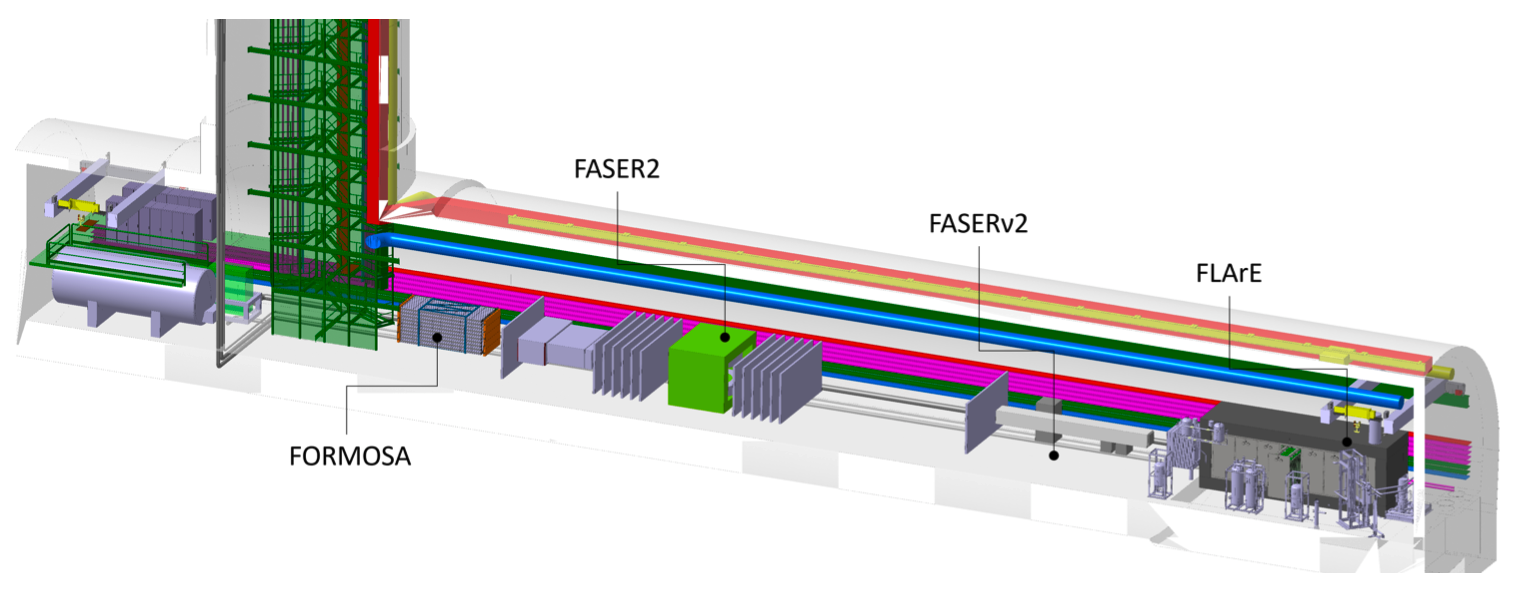}
\vspace{-2em}

\caption{\label{fig:ExecutiveSummaryMap}The proposed baseline layout of the Forward Physics Facility, located on the collision axis line-of-sight of the ATLAS experiment. The FPF will be $75\,\rm{m}$ long and $11.8\,\rm{m}$ wide and will house a diverse set of experiments to explore the many physics opportunities in the far-forward region.}
\end{figure}

Currently, plans are underway to host four experiments at the FPF, each utilizing distinct detector technologies tailored to specific physics objectives\,\cite{FPF_WP}. The proposed experiments are shown in the baseline layout in \cref{fig:ExecutiveSummaryMap}:
\begin{itemize}
\vspace{-0.2em}
\setlength{\itemsep}{2pt}
\setlength{\parskip}{0pt}
  \setlength{\parsep}{0pt}
\item {\bf FASER2:} A magnetic tracking spectrometer, designed to search for light and weakly-interacting states, including new force carriers, sterile neutrinos, axion-like particles, among others, and to distinguish $\nu$ and $\bar\nu$ charged current scattering in the upstream detectors.
\item {\bf FASER$\nu$2:} An on-axis emulsion detector, with pseudorapidity range $\eta>8.4$, that will detect TeV neutrinos with unparalleled spatial resolution, including tau neutrinos.
\item {\bf FLArE:} A noble liquid fine-grained time projection chamber to detect neutrinos and search for light dark matter with high kinematic resolution and wide dynamic range.
\item {\bf FORMOSA:} A detector composed of scintillating bars, with world-leading sensitivity to millicharged particles across a large range of masses.
\end{itemize}

These experiments will explore a wide range of physics phenomena\,\cite{FPF_SP,FPF_WP}, such as new particles, neutrinos, dark matter, dark sectors, and QCD. In this article, we will emphasize the unique interdisciplinary opportunities at the FPF for studies at the intersection of high-energy particle physics and astroparticle physics.

%------------------------
%------------------------

\section{Astroparticle Physics at the FPF}

Measurements conducted at the FPF will probe high-energy hadronic interactions in the far-forward region. These measurements aim to enhance the modeling of such interactions in the atmosphere, thereby reducing uncertainties in air shower observations and contributing to a better understanding of the properties of the highest-energy cosmic rays\,\cite{Kampert,MuonPuzzle}. Moreover, atmospheric neutrinos generated in extensive air showers in the far-forward region constitute a significant background for searches for high-energy astrophysical neutrinos using large-scale neutrino telescopes\,\cite{IceCube,IceCube_Gen2,KM3Net}. Therefore, measurements at the FPF will play a crucial role in refining our understanding of the atmospheric neutrino flux and minimizing associated uncertainties in searches for high-energy astrophysical neutrinos. The direct connections between measurements at the FPF and astroparticle physics will be further explored in the following.

\subsection{Light Hadron Production}
\label{sec:light_hadrons}

Muons serve as tracers of hadronic interactions, making their measurement in extensive air showers crucial for testing hadronic interaction models. Over the past two decades, various EAS experiments have reported discrepancies between model predictions and experimental data, referred to as the muon puzzle in EASs\,\cite{MuonPuzzle}. In particular, evidence for a deficit in muon numbers compared to simulations has been observed in analyses from the Pierre Auger Observatory\,\cite{Auger1,Auger2} and a systematic meta-analysis of data from nine air shower experiments revealed an energy-dependent trend of these discrepancies with high statistical significance\,\cite{WHISP1,WHISP2,WHISP3}. Recent studies suggest that these observed discrepancies may have a very complex nature\,\cite{WHISP4} and indicate severe deficits in our understanding of particle physics which are currently not understood\,\cite{MuonPuzzle}.

Muons detected in air shower experiments range typically from a few to tens of GeV. They are produced at the end of the cascade of hadronic interactions spanning up to approximately ten generations, predominantly through soft hadron production processes. \Cref{fig:eta-ranges} shows that hadron production at forward pseudorapidities has the largest influence on muon generation in EASs. The sensitivity to these hadrons is substantial, where even small deviations of around $5\%$ in their multiplicity or type can have sizable impact the resulting muon flux\,\cite{Baur}.

%------------------------
\begin{figure}[tb]
\centering
\vspace{-1em}
\includegraphics[width=.95\textwidth]{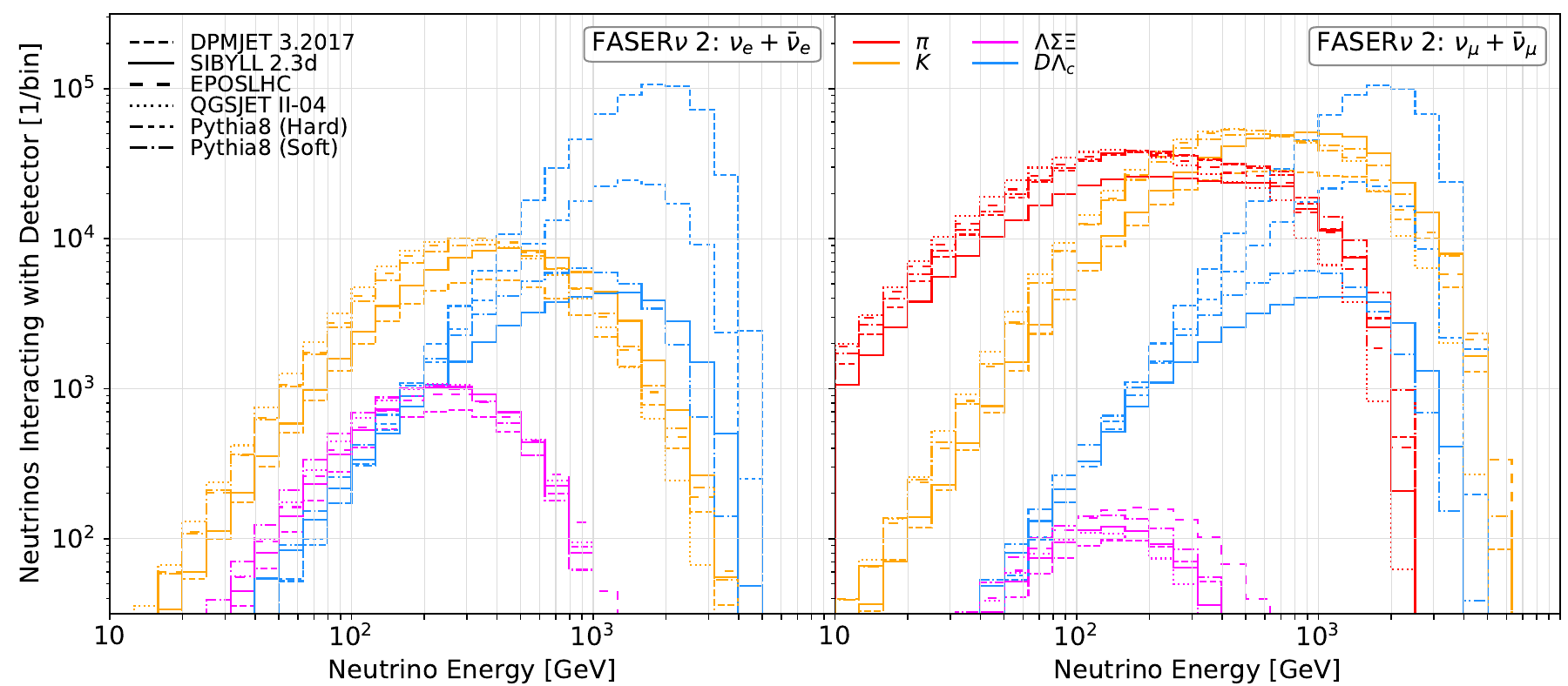}
\caption{\label{fig:nu-rate}Neutrino energy spectra\,\protect\cite{FPF_WP,Kling} for electron neutrinos (left) and muon neutrinos (right) passing through FASER$\nu$2 for an integrated luminosity of $3\,\rm{ab}^{-1}$. The different production modes are shown separately, i.e., pion decays (red), kaon decays (orange), hyperon decays (magenta), and charm decays (blue). The predictions are obtained from SIBYLL-2.3d (solid), DPMJET-III.2017 (short dashed), EPOS-LHC (long dashed), QGSJET-II.04 (dotted), and Pythia~8.2 using soft-QCD processes (dot-dashed) and with hard-QCD processes for charm production (double-dot-dashed).}
\vspace{-1em}
\end{figure}
%------------------------

Dedicated measurements at the FPF will significantly enhance our understanding of light hadron production in the far-forward region. The ratio of electron neutrino ($\nu_e$) to muon neutrino~($\nu_\mu$) fluxes measured in FPF experiments serves as an indirect measure of the ratio of charged kaons ($K$) to pions ($\pi$). Electron neutrino fluxes are predominantly produced from kaons, while muon neutrinos originate from both pion and kaon decays. However, $\nu_e$ and $\nu_\mu$ have distinct energy spectra, enabling their differentiation. In addition, neutrinos from pion decay are more concentrated around the LOS than those from kaons, given that $m_\pi < m_K$, and thus neutrinos from pions obtain less additional transverse momentum than those from kaon decays. Thereby, the closeness of the neutrinos to the LOS, or equivalently their rapidity distribution, can be used to disentangle different neutrino origins to get an estimate of the pion-to-kaon ratio. \Cref{fig:nu-rate} shows predictions of the neutrino energy spectra for $\nu_e$ and $\nu_\mu$ interacting in FASER$\nu$2, assuming an integrated luminosity of $3\,\rm{ab}^{-1}$. These predictions are derived from various models, including SIBYLL-2.3d\,\cite{SIBYLL}, DPMJET-III\,\cite{DPMJET}, EPOS-LHC\,\cite{EPOS}, QGSJET-II.04\,\cite{QGSJET}, and Pythia~8\,\cite{PYTHIA}. The differences in the resulting fluxes exceed a factor of two, which is much larger than the expected statistical uncertainties at the FPF\,\cite{Kling}. Since the muon puzzle is assumed to be of soft-QCD origin, there is also a strong connection to the QCD program of the FPF and dedicated QCD measurements\,\cite{FPF_WP} will further help to understand particle production in EAS.

%------------------------
\begin{figure}[tb]
\centering
\vspace{-1em}
\includegraphics[width=1.\textwidth]{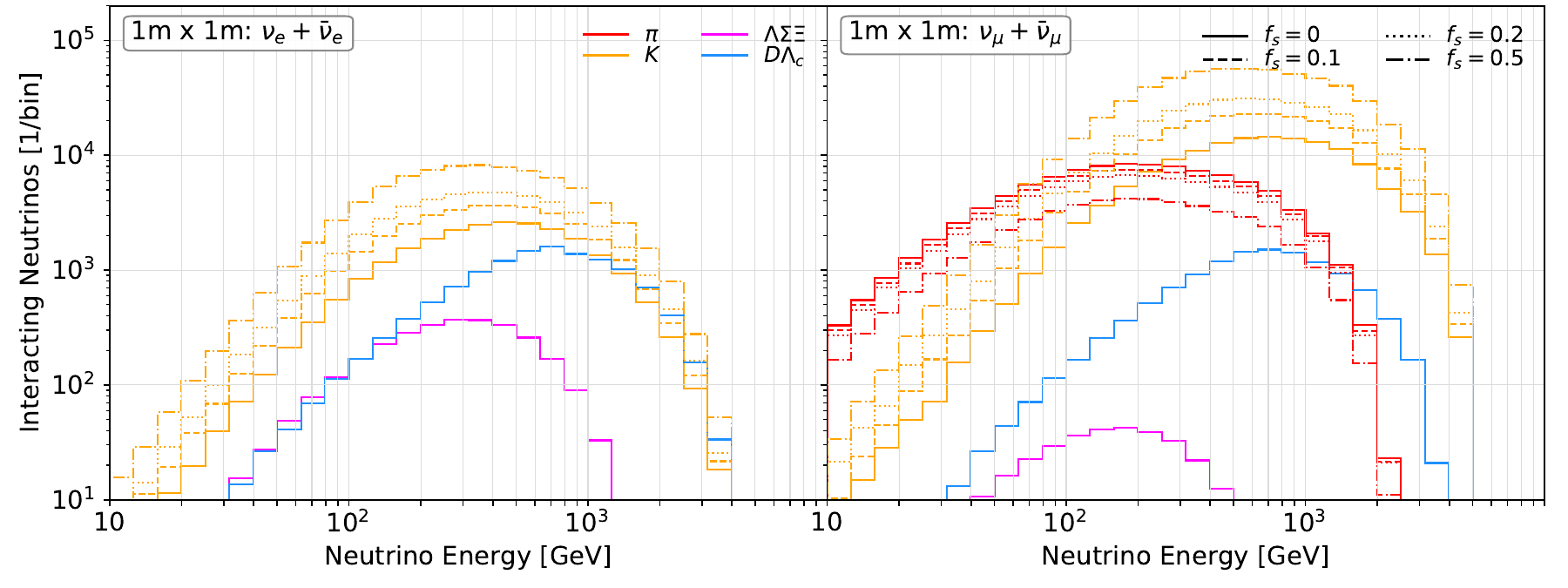}
\vspace{-1.2em}
\caption{\label{fig:dos}Neutrino energy spectra\,\protect\cite{Anchordoqui,Sciutto} for electron neutrinos (left) and muon neutrinos (right) passing through the FLArE detector (10-ton target). The vertical axis shows the number of neutrinos that pass the detector's cross-sectional area of $1\,\mathrm{m}^2$ for an integrated luminosity of $3\,{\rm ab}^{-1}$: pion decays (red), kaon decays (orange), hyperon decays (magenta), and charm decays (blue). The different line styles correspond to predictions obtained from SIBYLL-2.3d by varying $f_s$.}
\end{figure}
%------------------------

A potential key to understanding the muon puzzle is the universal strangeness and baryon enhancement observed by ALICE in collisions at mid-rapidity\,\cite{ALICE}. This enhancement depends solely on the particle multiplicity and not on the details of the collision system, allowing predictions of hadron composition in EAS in a phase space beyond current collider capabilities. If the enhancement increases in the forward region, it will impact muon production in EASs and could be traced by the ratio of charged kaons to pions measured at the FPF. \Cref{fig:dos} shows predictions from a simple toy model\,\cite{Anchordoqui,Sciutto} based on SIBYLL-2.3d, where strangeness enhancement is introduced by allowing to swap $\pi \rightarrow K$ with a probability $f_s$ at large pseudorapidities ($\eta>4$). For $f_s = 0.1$ ($f_s = 0.2$), the predicted electron neutrino flux is a factor of 1.6 (2.2) higher at its maximum than the baseline prediction. These differences are significantly larger than the anticipated uncertainties at the FPF. It has been shown that for $0.4 < f_s < 0.6$ this simple model can partially accommodate data from the Pierre Auger Observatory \cite{Anchordoqui}, thus providing a potential explanation for the muon puzzle. This example demonstrates how the FPF will be able to uniquely test and constrain hadronic interaction models, fundamentally improving our understanding of multi-particle production in EASs.

In addition, it might also be possible to use forward-going muons to constrain the production of pions and kaons. The muon flux at FASER2, for example, is estimated to be approximately $1\,\rm{kHz}$ per $\rm{cm}^2$. Studies of these muons could provide complementary information to determine the ratio of charged pions to kaons. However, these measurements may be challenging because the origin of the muon flux at the FPF is not well understood and further investigations based on detailed simulations are currently ongoing.

\subsection{Charm Hadron Production}

High-energy neutrinos of astrophysical origin are routinely observed by large-scale neutrino telescopes. The upcoming generation of telescopes, such as IceCube-Gen2\,\cite{IceCube_Gen2} and KM3NeT\,\cite{KM3Net}, are expected to detect significantly more astrophysical neutrinos. However, atmospheric neutrinos produced in EASs in the atmosphere present an irreducible background for these searches. Completely eliminating atmospheric background events is experimentally not feasible, instead, these backgrounds must be estimated and subtracted. The acceptance for these events is usually determined through simulations, highlighting the need for precise experimental data to test and refine these simulations, thereby reducing associated uncertainties. Accurate understanding of cosmic sources thus requires a comprehensive understanding of neutrino production in high-energy hadron interactions, particularly from heavy hadron decays.

Atmospheric neutrinos are generated by the semileptonic decays of hadrons in EAS, typically from $\pi$ and $K$ decays (see also \cref{sec:light_hadrons}). This is known as the conventional neutrino flux, which decreases with increasing energy. At sufficiently high energies, however, atmospheric neutrinos are also generated from the semileptonic decay of heavy flavor hadrons, such as $D$ mesons, $B$ mesons, and $\Lambda_c$ baryons, referred to as the prompt neutrino flux. Due to their very short decay lengths, these hadrons decay immediately to neutrinos upon production, causing the prompt neutrino flux to decrease more slowly with energy than the conventional flux. At high energies ($E_\nu \sim 10^5-10^6 \,\rm{GeV}$), prompt atmospheric neutrinos become the main background for astrophysical neutrino searches. For neutrino energies of $10^6\,\rm{GeV}$, the corresponding center-of-mass energy is around $8\,\rm{TeV}$, a range that can be explored at the HL-LHC.

\begin{figure}[tb]
\vspace{-1em}
\mbox{
\includegraphics[width=.49\linewidth]{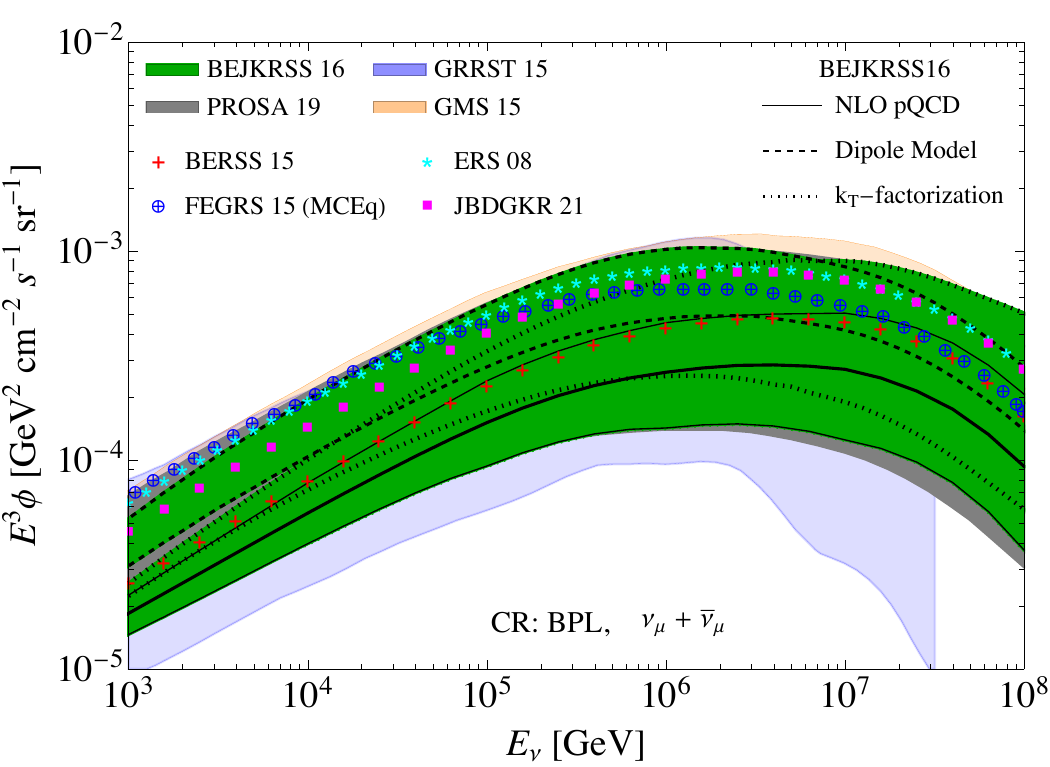}\,
\includegraphics[width=.49\linewidth]{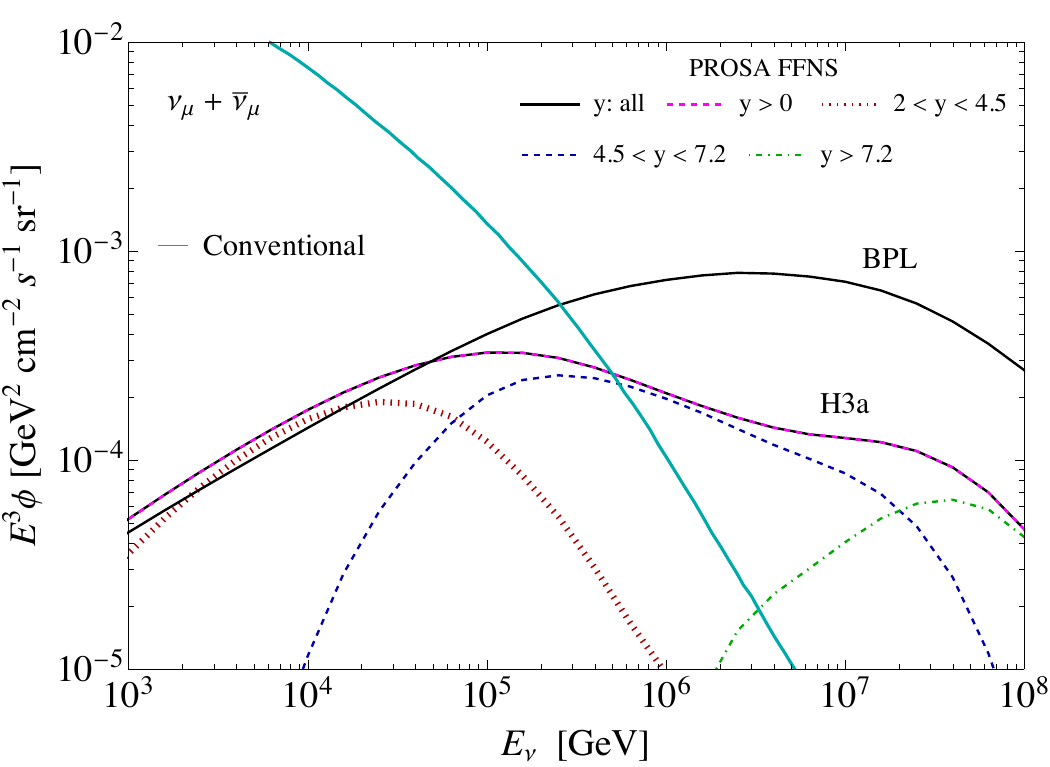}
}
\caption{\label{fig:PromptAtmNu}Comparison of the prompt atmospheric muon neutrino flux from various recent flux calculations\,\protect\cite{PROSA} (left). The incident cosmic-ray flux is approximated with a broken power law in all predictions. If available, the accociated uncertainties are shown as error band. Also shown is the prompt atmospheric muon neutrino flux produced in different collider rapidity ranges\,\protect\cite{Jeong2} (right) and a prediction of the conventional atmospheric neutrino flux\,\protect\cite{Honda}. The predictions are shown assuming a simple broken power law (BPL) and the Gaisser H3a flux model\,\protect\cite{H3a} for the initial cosmic-ray flux.
}  
\end{figure}

While \cref{fig:nu-rate,fig:dos} provide predictions of the charm contribution to the neutrino flux at the FPF based on various hadronic interaction models, \cref{fig:PromptAtmNu} shows several predictions for the prompt atmospheric muon neutrino flux based on pQCD calculations\,\cite{PROSA}, assuming a simple broken power law (BPL) cosmic-ray spectrum. The left panel of \cref{fig:PromptAtmNu} shows that theoretical predictions of the prompt atmospheric neutrino flux have substantial uncertainties. These uncertainties arise from various input parameters in the calculation of prompt fluxes, such as cross-sections for heavy flavor production, parton distribution functions, and fragmentation functions. The right panel shows that an additional significant uncertainty originates from the cosmic-ray flux assumption. Moreover, \cref{fig:PromptAtmNu} (right) illustrates the individual contributions from charm hadrons produced at different rapidity ranges. At energies where the prompt component dominates, the prompt atmospheric neutrino flux originates from charm hadrons produced at rapidities of $y > 4.5$, accessible by the proposed FPF experiments at the HL-LHC\,\cite{Jeong1}.

%------------------------
%------------------------

The FPF will provide data with unprecedented statistics in the far-forward region, at energies that are relevant to astrophysical neutrino searches. This will provide crucial information to reduce uncertainties and strongly constrain models of charm hadron production. Thus, the FPF will be pivotal in improving predictions of the prompt atmospheric neutrino flux, which will significantly impact searches for astrophysical neutrinos by reducing the associated uncertainties.

\section{Conclusions}

The Forward Physics Facility is a proposal to construct a new underground cavern at the high-luminosity LHC, which will host a variety of experiments focused on particle production in the far-forward region. These experiments will enable unique neutrino measurements with unprecedented statistics, probing hadron production in a phase space that is inaccessible to any existing collider experiment but crucial for modeling high-energy hadronic interactions in the atmosphere. Consequently, the measurements at the FPF are closely linked to open questions in modern astroparticle physics. They will provide unique tests to constrain hadronic interaction models, fundamentally improving our understanding of multi-particle production in extensive air showers, and thereby notably contribute to solving the muon puzzle. This will significantly reduce the uncertainties in ground-based cosmic-ray observations which typically rely on interpretations based on these models. Additionally, measurements at the FPF will provide crucial information about charm hadron production in the forward region which will improve predictions of the prompt atmospheric neutrino flux and reduce associated uncertainties in searches for astrophysical neutrinos. Thus, the experiments at the FPF will contribute to enhance our understanding of the origin and nature of the highest energy cosmic rays and astrophysical neutrinos in the context of future multi-messenger observations\,\cite{ASTRO2020,Coleman}.


\vspace{2em}


\section*{References}

\begin{thebibliography}{99}


\bibitem{Kampert}K.-H. Kampert and M.~Unger, \Journal{\em Astropart. Phys.}{35}{660-678}{2012}.

\bibitem{Engel}R. Engel, D. Heck, and T. Pierog, \Journal{\em Ann. Rev. Nucl. Part. Sci.}{61}{467-489}{2011}.

\bibitem{MuonPuzzle}J.~Albrecht {\it et al.}, \Journal{\em Astrophys. Space Sci.}{367}{27}{2022}.

\bibitem{FPF_SP}L.~A.~Anchordoqui {\it et al.}, \Journal{\PRepD}{968}{1-50}{2022}.

\bibitem{FPF_WP}J.~L.~Feng {\it et al.}, \Journal{\JPG}{50}{030501}{2023}.

\bibitem{FASER1}H.~Abreu {\it et al.}, \Journal{\PRL}{131}{031801}{2023}.

\bibitem{SNDLHC1}R.~Albanese {\it et al.}, \Journal{\PRL}{131}{031802}{2023}.

\bibitem{Auger1}A.~Aab {\it et al.} (Pierre Auger Collaboration), \Journal{\PRD}{91}{032003}{2015}.

\bibitem{Auger2}A.~Aab {\it et al.} (Pierre Auger Collaboration), \Journal{\PRL}{117}{192001}{2016}.

\bibitem{WHISP1}H.~P.~Dembinski {\it et al.} (EAS-MSU, IceCube, KASCADE Grande, NEVOD-DECOR, Pierre Auger, SUGAR, Telescope Array, Yakutsk EAS Array Collaborations), \Journal{\EPJConf}{210}{02004}{2019}.

\bibitem{WHISP2}L.~Cazon (EAS-MSU, IceCube, KASCADE Grande, NEVOD-DECOR, Pierre Auger, SUGAR, Telescope Array, Yakutsk EAS Array Collaborations), \Journal{\em PoS}{ICRC2019}{214}{2019}.

\bibitem{WHISP3}D.~Soldin (EAS-MSU, IceCube, KASCADE Grande, NEVOD-DECOR, Pierre Auger, SUGAR, Telescope Array, Yakutsk EAS Array Collaborations), \Journal{\em PoS}{ICRC2021}{349}{2021}.

\bibitem{WHISP4}J.~C.~Arteaga~Velazquez, \Journal{\em PoS}{ICRC2023}{466}{2023}.

\bibitem{Baur}S.~Baur {\it et al.}, \Journal{\PRD}{107}{094031}{2023}.

\bibitem{Kling}F. Kling and L. J. Nevay \Journal{\PRD}{104}{113008}{2021}.

\bibitem{SIBYLL}F.~Riehn {\it et al.}, \Journal{\PRD}{102}{063002}{2020}.

\bibitem{DPMJET}J.~Ranft, R.~Engel, and S.~Roesler {\it et al.}, \Journal{\em Nucl. Phys. B Proc. Suppl.}{122}{392-395}{2003}.

\bibitem{EPOS}T.~Pierog {\it et al.}, \Journal{\PRC}{92}{034906}{2015}.

\bibitem{QGSJET}S.~Ostapchenko, \Journal{\EPJConf}{208}{11001}{2019}.

\bibitem{PYTHIA}T.~Sjöstrand {\it et al.}, \Journal{\em Comput. Phys. Commun.}{191}{159-177}{2015}.

\bibitem{ALICE}J.~Adam {\it et al.} (ALICE Collaboration), \Journal{\em Nature Phys.}{13}{353-539}{2017}.

\bibitem{Anchordoqui}L.~A.~Anchordoqui {\it et al.}, \Journal{\em JHEAP}{34}{19-32}{2022}.

\bibitem{Sciutto}S.~J.~Sciutto {\it et al.}, \Journal{\em PoS}{ICRC2023}{388}{2023}.

\bibitem{IceCube}M.~G.~Aartsen {\it et al.} (IceCube Collaboration), \Journal{\em JINST}{12}{P03012}{2017}.

\bibitem{IceCube_Gen2}M.~G.~Aartsen {\it et al.} (IceCube-Gen2 Collaboration), \Journal{\JPG}{48}{060501}{2021}.

\bibitem{KM3Net}R.~Coniglione {\it et al.} (KM3NeT Collaboration), \Journal{\em JPCS}{632}{012002}{2015}.

\bibitem{PROSA}O.~Zenaiev {\it et al.}, \Journal{\em JHEP}{04}{118}{2020}.

\bibitem{Jeong1}Y.~S.~Jeong {\it et al.}, \Journal{\em PoS}{ICRC2023}{968}{2023}.

\bibitem{Jeong2}Y.~S.~Jeong {\it et al.}, \Journal{\em PoS}{ICRC2021}{1218}{2021}.

\bibitem{Honda}M.~Honda {\it et al.}, \Journal{\PRD}{75}{043006}{2007}.

\bibitem{H3a}T.~K.~Gaisser {\it et al.}, \Journal{\em Astropart. Phys.}{35}{801-806}{2012}.

\bibitem{ASTRO2020}F.~G.~Schröder {\it et al.}, \Journal{\em BAAS}{51}{131}{2019}.

\bibitem{Coleman}A.~Coleman {\it et al.}, \Journal{\em Astropart. Phys.}{149}{102819}{2023}.




\end{thebibliography}
\end{document}